\documentclass[prl, preprintnumbers ,showpacs,amsmath,amssymb, superscriptaddress,twocolumn]{revtex4-1}
\usepackage{bm}
\usepackage{amsmath}
\usepackage{amssymb}
\usepackage{amsthm}
\usepackage{amsfonts}
\usepackage{enumerate}
\usepackage{latexsym}
\usepackage{times}
\usepackage{ulem}
\usepackage{graphicx}
\usepackage{color}
\usepackage{makeidx}
\usepackage{ifpdf}

\begin{document}

\title{Strain-induced nonsymmorphic symmetry breaking and removal of Dirac semimetallic nodal line in an orthoperovskite iridate}

\author{Jian~Liu} \email{jianliu@utk.edu}
\affiliation{Department of Physics and Astronomy, University of Tennessee, Knoxville, Tennessee 37996, USA}
\affiliation{Department of Physics, University of California, Berkeley, California 94720, USA}
\affiliation{Materials Science Division, Lawrence Berkeley National Laboratory, Berkeley, California 94720, USA}

\author{D. Kriegner} \email{dominik.kriegner@gmail.com}
\author{L. Horak}
\affiliation{Faculty of Mathematics and Physics, Charles University, Ke Karlovu 5, 121 16 Prague, Czech Republic}
\author{D. Puggioni}
\affiliation{Department of Materials Science and Engineering, Northwestern University, Evanston, Illinois 60208, USA}
\author{C.~Rayan~Serrao}
\affiliation{Department of Physics, University of California, Berkeley, California 94720, USA}
\affiliation{Department of Materials Science and Engineering, University of California, Berkeley, California 94720, USA}
\author{R.~Chen}
\affiliation{Department of Physics, University of California, Berkeley, California 94720, USA}
\affiliation{Materials Science Division, Lawrence Berkeley National Laboratory, Berkeley, California 94720, USA}
\author{D.~Yi}
\affiliation{Department of Materials Science and Engineering, University of California, Berkeley, California 94720, USA}
\author{C. Frontera}
\affiliation{Institut de Ci\`{e}ncia de Materials de Barcelona (ICMAB-CSIC), Campus de la UAB, E-08193 Bellaterra, Spain}
\author{V. Holy}
\affiliation{Faculty of Mathematics and Physics, Charles University, Ke Karlovu 5, 121 16 Prague, Czech Republic}
\author{A.~Vishwanath}
\affiliation{Department of Physics, University of California, Berkeley, California 94720, USA}
\affiliation{Materials Science Division, Lawrence Berkeley National Laboratory, Berkeley, California 94720, USA}
\author{J. M.\ Rondinelli}
\affiliation{Department of Materials Science and Engineering, Northwestern University, Evanston, Illinois 60208, USA}
\author{X.~Marti}
\affiliation{Department of Spintronics and Nanoelectronics, Institute of Physics ASCR, v.v.i., Cukrovarnick\'{a} 10, 162 53 Praha 6, Czech Republic}
\author{R. Ramesh}
\affiliation{Department of Physics, University of California, Berkeley, California 94720, USA}
\affiliation{Materials Science Division, Lawrence Berkeley National Laboratory, Berkeley, California 94720, USA}
\affiliation{Department of Materials Science and Engineering, University of California, Berkeley, California 94720, USA}


\begin{abstract}
   By using a combination of heteroepitaxial growth, structure refinement based on synchrotron x-ray diffraction and first-principles calculations, we show that the symmetry-protected Dirac line nodes in the topological semimetallic perovskite SrIrO$_3$ can be lifted simply by applying epitaxial constraints. In particular, the Dirac gap opens without breaking the $Pbnm$ mirror symmetry. In virtue of a symmetry-breaking analysis, we demonstrate that the original symmetry protection is related to the $n$-glide operation, which can be selectively broken by different heteroepitaxial structures. This symmetry protection renders the nodal line a nonsymmorphic Dirac semimetallic state. The results highlight the vital role of crystal symmetry in spin-orbit-coupled correlated oxides and provide a foundation for experimental realization of topological insulators in iridate-based heterostructures.
\end{abstract}

\maketitle
\section{I. Introduction}

Since the discovery of topological states in semiconductors, research on materials with strong relativistic coupling of electron spin and orbital momenta, the spin-orbit coupling (SOC), have been flourishing owing to the huge potential for such materials to support spin-based electronics \cite{Hasan2010}. One of the most important frontiers focuses on the integration of SOC into the rich many-body physics of correlated electron systems, in the search for novel phases, for example, a topological Mott insulator, Weyl semimetal, quantum anomalous Hall effect, spin liquid, and unconventional superconductivity in transition metal oxides \cite{Pesin2010,Xiao2011,Ran2009,Wan2011,Wang2011}.

As a model system, perovskite SrIrO$_3$ plays a key role in both addressing the SOC-correlation interplay and searching for novel topological phases in correlated oxides. Specifically, while the SOC-enhanced Mott instability has been shown to induce an unusual insulating state in Sr$_{2}$IrO$_{4}$ and Sr$_{3}$Ir$_{2}$O$_{7}$ from the Ruddlesden-Popper series Sr$_{n+1}$Ir$_{n}$O$_{3n+1}$ \cite{Kim2008,Moon2008}, the $n=\infty$ member SrIrO$_{3}$ remains metallic \cite{Longo1971,Zhao2008,Cheng2011}. This dimensionality-controlled insulator-to-metal transition was believed to be caused by increased bandwidth \cite{Moon2008}. But recent angle-resolved photoemission spectroscopy (ARPES) \cite{Nie2015} indicated that the bandwidth of SrIrO$_{3}$ is even narrower than Sr$_{2}$IrO$_{4}$ with a semimetallic Fermi surface. Band structure calculations \cite{Carter2012,Zeb2012} have showed that, although SOC does aid in reducing the density of states at the Fermi level, full gap opening is prevented by band crossings at a Dirac nodal ring in the $U-R-X$ plane originating from the mirror reflection of the crystalline $Pbnm$ symmetry, rendering a topological semimetallic state. Moreover, it was suggested that lifting the Dirac degeneracy by breaking the mirror symmetry would lead to various topological surface states in different heteroepitaxial superlattices \cite{Carter2012,Chen2014,Chen2015}. Experimentally, an ARPES study on epitaxial thin film samples observed a gap between Dirac bands rather than a nodal ring \cite{Liu2015}. It is, however, unclear how the symmetry protection and electronic structure would respond to epitaxial strain, which is essential prior to the realization of heterostructures. In fact, because the perovskite polymorph is not the thermodynamically stable phase of SrIrO$_{3}$ and bulk synthesis yields only polycrystalline samples \cite{Longo1971,Zhao2008,Cheng2011}, epitaxial thin films are featured in the investigations of SrIrO$_3$ \cite{Moon2008,Wu2010,Biswas2008,Gruenewald2014,Zhang2015}, including the two ARPES studies \cite{Nie2015,Liu2015} mentioned above, highlighting the necessity for understanding the influence of epitaxy on the symmetry.

Here, using heteroepitaxial growth, structure refinement based on synchrotron x-ray diffraction and first-principles calculations, we show that an epitaxial constraint imposed by the substrate breaks the symmetry that protects the Dirac line nodes, and opens a Dirac gap. In particular, the results reveal that the Dirac degeneracy is lifted without breaking the mirror symmetry, which is contrary to the previous theoretical proposals and indicates alternative possibilities for stabilizing topological states. A symmetry comparison further shows that another symmetry operation, i.e. the $n$-glide plane, within the $Pbnm$ space group plays a more fundamental role in protecting the Dirac line nodes in orthoperovskite iridates, placing the system's topological band crossing in the category of a three-dimensional nonsymmorphic Dirac semimetal.

\section{II. Experimental Details}
\begin{figure}\vspace{-0pt}
\includegraphics[width=8cm]{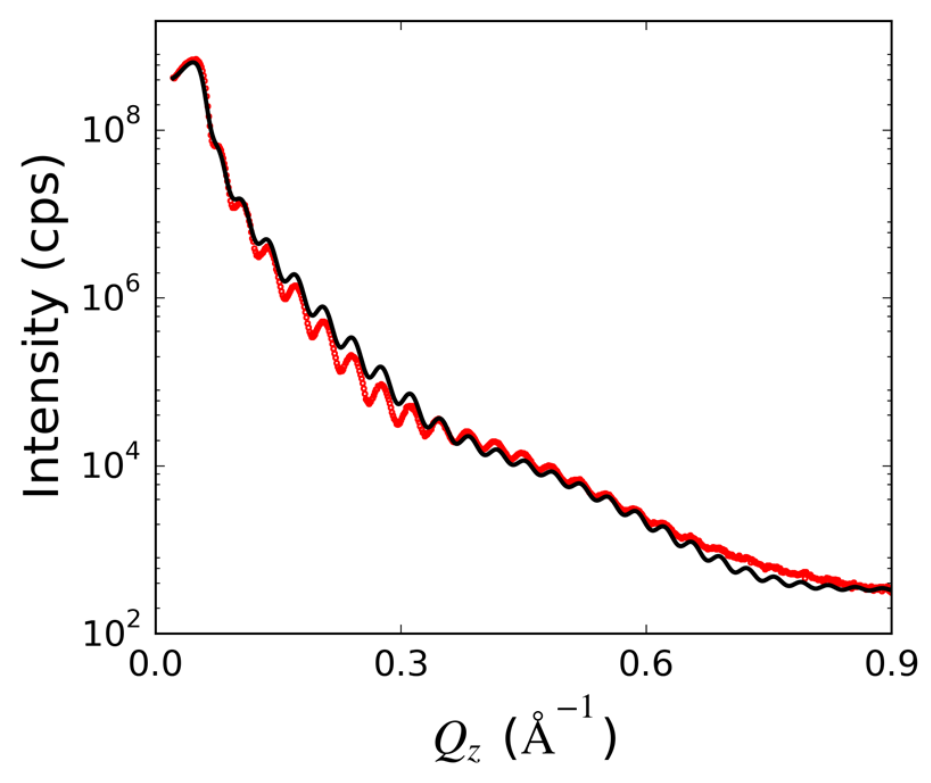}
\caption{\label{XRR} X-ray reflectivity data (red) and simulation (black). Thickness fringes and the reflectivity signal extend to high-momentum transfers, indicating the low roughness of the film/substrate interface and surface. Simulations with the Parratt formalism \cite{Holy} yield a film thickness of 18 nm and a rms roughness between 0.2 and 0.4 nm for the substrate/film interface as well as the surface.
}
\end{figure}

High-quality single-crystal thin films ($\sim$18 nm) of perovskite SrIrO$_{3}$ were grown by pulsed laser deposition \cite{Serrao2013,Jian2013} on (110)-oriented GdScO$_3$ ($Pbnm$), which applies $\sim0.6\%$ tensile strain.
In-house and synchrotron-based x-ray diffraction experiments at beamline 6IDB of the Advanced Photon Source confirmed the highly epitaxial single-phase perovskite structure and the film thickness (Fig.~\ref{XRR}). X-ray diffraction experiments along the truncation rods under grazing-incidence geometry were performed at SpLine (BM25, European Synchrotron Radiation Facility) with a fixed incidence angle of 1$^\circ$ degree and an x-ray photon energy of 15 keV. Using this geometry, we performed crystal truncation-rod-scans along the surface normal ([110] direction) for more than 70 Bragg peaks.

%

\section{III. Results and Analysis}
\begin{figure*}\vspace{-0pt}
\includegraphics[width=16cm]{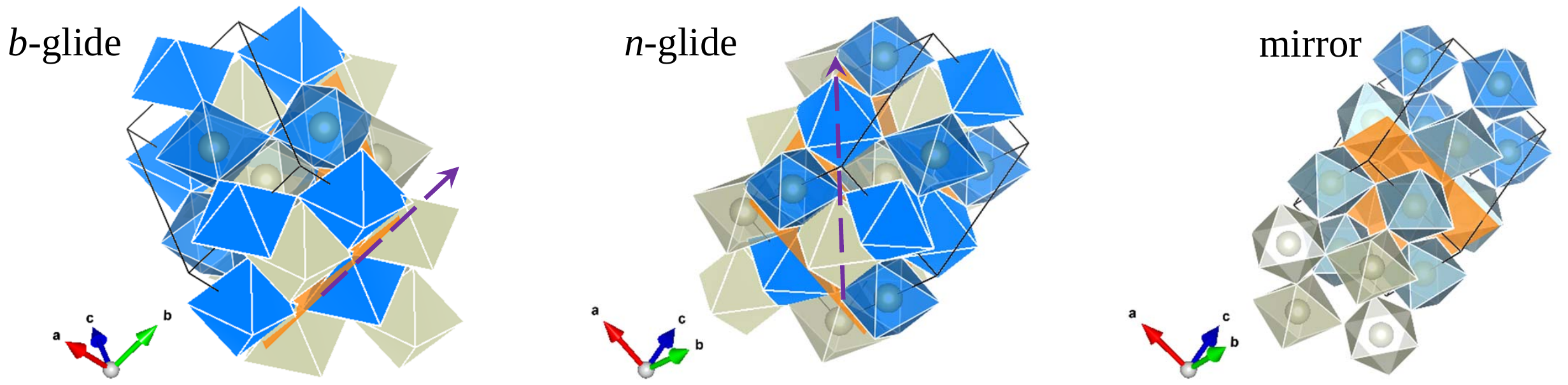}
\caption{\label{symmetry} Schematics for the three symmetry operations of the $Pbnm$ perovskite $AB$O$_{3}$ structure. For simplicity, only the $B$O$_{6}$ octahedra are shown. The blue octahedra with the black box denote the original $Pbnm$ unit cell. The gray octahedra denote the unit cell after the reflection operation where the reflection plane is highlighted in orange. For the $b$-glide and $n$-glide symmetries, the solid/transparent octahedra denote the equivalent sites under each symmetry, respectively. The glide directions are denoted by dashed purple arrows. Since the mirror operation generates the enantiomorphic pair, the octahedra adjacent to the reflection plane become blue-gray due to superposition of the original and mirrored unit cells.
}
\end{figure*}

\textit{Symmetry}. It is helpful to first illustrate the lattice symmetry operations of perovskite SrIrO$_3$, which has the orthorhombic GdFeO$_{3}$-type structure owing to the $a^{-}a^{-}c^{+}$ octahedral rotation \cite{Glazer1975}, the same as the GdScO$_{3}$ substrate. This creates a four-formula unit cell from a $\sqrt{2}\times\sqrt{2}\times2$ cell multiplication of the cubic perovskite cell. The [110] growth direction is thus along a pseudo-cubic [100]$_{\rm c}$ direction. The $Pbnm$ space group, shorthand for $P\,2_{1}/b\,2_{1}/n\,2_{1}/m$, has a $b$-glide plane perpendicular to $a$ (a $bc$ plane reflection plus a translation along $b$), an $n$-glide plane perpendicular to $b$ (an $ac$ plane reflection plus a diagonal translation within the $ac$ plane), and a mirror plane perpendicular to $c$, as shown in Fig.~\ref{symmetry}. The $b$-glide and $n$-glide operations are nonsymmorphic symmetries, rendering $Pbnm$ a nonsymmorphic space group. Tight-binding calculations suggested a Dirac nodal ring around the $U$ point in the $X-U-R$ plane originating from the mirror symmetry \cite{Carter2012}. In the following, we will experimentally show that this symmetry is preserved but the other two are broken in the strained film.

\textit{X-ray diffraction and monoclinicity}. In order to determine the unit-cell parameters, symmetry, and atomic positions of the SrIrO$_3$ thin film, $>$70 different Bragg peaks were measured in surface x-ray diffraction geometry (fixed incidence angle) using truncation rod scans along the surface normal [110] direction. The film diffraction signal is always found to have the same in-plane momentum transfer as the substrate, demonstrating the fully strained epitaxial growth. We also found intensity from the film near the (221) position of the substrate (Fig.\ref{XRD}b), indicating the bulk-like $a^{-}a^{-}c^{+}$ octahedra rotation (in fact $a^{-}b^{-}c^{+}$ as shown below) and discarding the other two; the [001] and [1$\overline{1}$0] axes of the film are locked to that of the substrate within the surface plane and not rotated 90$^\circ$.

\begin{figure*}\vspace{-0pt}
\includegraphics[width=18cm]{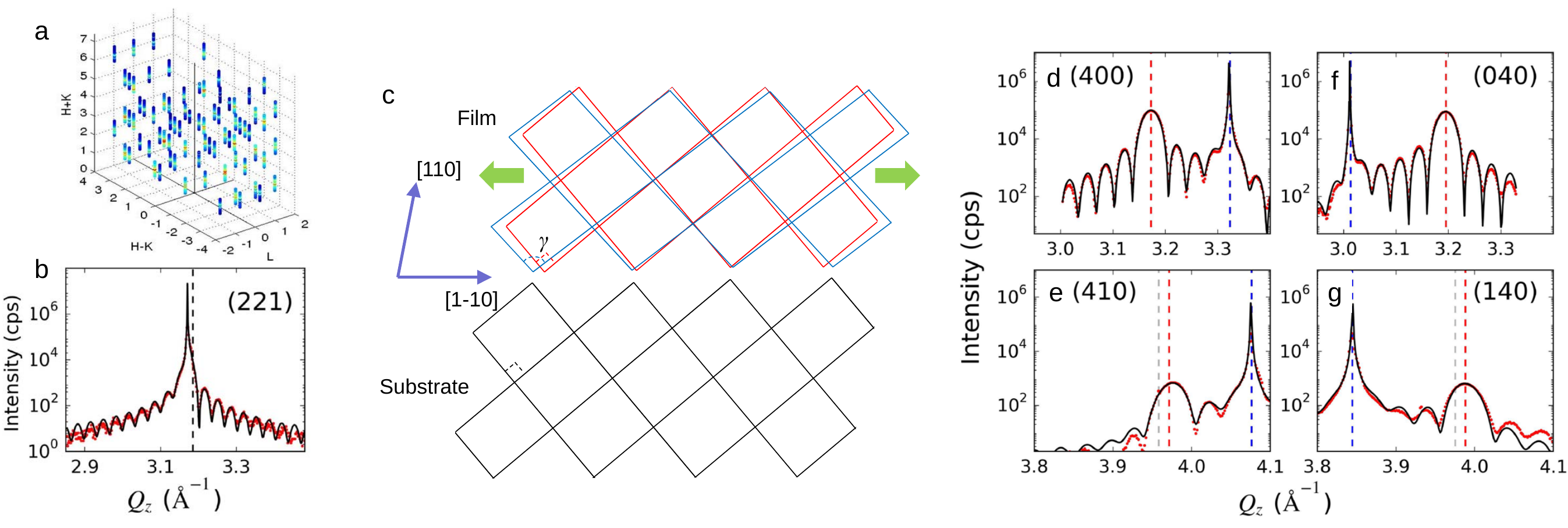}
\caption{\label{XRD} a) The scattered intensity along the truncation rods in over 70 reflections obtained under grazing incidence diffraction geometry and plotted against the reciprocal lattice units of the GdScO$_3$ substrate. b) Truncation rod scan at the (221) Bragg peak. (c) The strain-induced monoclinicity between two orthorhombic lattice from the side view. Black, red, and blue show the substrate, unstrained and strained films, respectively. The orthorhombicity is exaggerated to illustrate the effect. Green arrows denote tensile strain. (d)-(g) Truncation rod scans at the (400), (040), (410), and (140) Bragg peaks, respectively. The diffraction signal from the substrate and thin film including Laue thickness fringes is experimentally observed (red) and reproduced by simulations (black). The peak position of the film and substrate are indicated as red and blue dashed lines, respectively. The gray dashed lines indicate the peak positions of (410) and (140) calculated based on that of (400) and (040) by fixing $\gamma=90^{\circ}$.
}
\end{figure*}

Although the condition of epitaxy itself does not fix any interaxial cell angles, we found that $\alpha$ and $\beta$ are $90^{\circ}$ within the error bars by analyzing the diffraction peak positions.
On the other hand, due to the lattice mismatch, the tensile strain introduces cell elongation along the [1$\overline{1}$0] direction and compression along the out-of-plane [110] direction (Fig.~\ref{XRD}c), leading to a deviation of $\gamma$ away from 90$^{o}$ and lowering the film lattice from $Pbnm$ presumably to a monoclinic subgroup. Indeed, this angular deviation is reflected in the relative peak positions within each $L$-plane of the reciprocal lattice. For instance, if one uses the (400) and (040) peak positions to calculate those of (410) and (140) by fixing $\gamma=90^{\circ}$, the simulated positions would clearly deviate away from that in the experimental scans and vice versa (Fig.~\ref{XRD}d-g). To account for this difference, $\gamma$ has to be increased by $\sim0.4^{\circ}$, consistent with the cell elongation along the [1$\bar{1}$0] direction.
Since $\gamma\neq90^{\circ}$, both the $b$- and $n$-glide operations become incompatible with the lattice and the most plausible space group is monoclinic $P112_{1}/m$ (no. 11), which in Glazer notation \cite{Glazer1975} corresponds to the $a^{-}b^{-}c^{+}$ pattern.

\begin{figure}\vspace{-0pt}
\includegraphics[width=8.5cm]{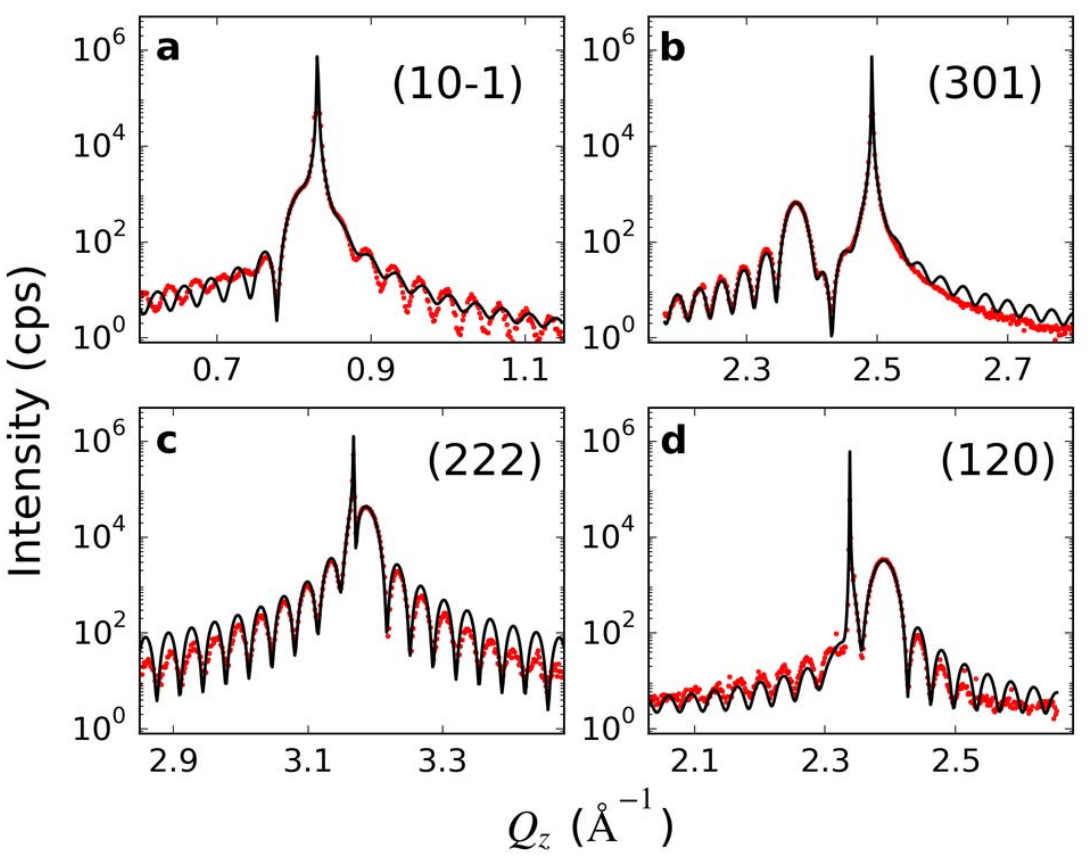}
\caption{\label{XRD_supl} Selected crystal truncation-rod-scans (red) compared with dynamical diffraction theory simulation (black). (a)-(d) The data and simulations of the Bragg peaks (10-1), (301), (222), and (120), respectively.
}
\end{figure}

To confirm this symmetry breaking, we simulated the complete set of truncation rods simultaneously and determined the structure factors. The simulation is required since a precise determination of the diffraction intensity and position of the SrIrO$_3$ layer is hindered by the interference from the diffracted wave of the substrate (Fig.~\ref{XRD}). This interference leads to a deformation of Laue fringes, especially when the film and substrate peaks are in close vicinity to each other (Fig.~\ref{XRD}b).
Using the thickness determined from x-ray reflectivity, we performed dynamical diffraction theory calculations \cite{Holy} in order to determine both the position and structure factor of the layer structure. These simulations nicely reproduce the measured diffractions curves, including the aforementioned distortions seen in Figs.~\ref{XRD} and \ref{XRD_supl}. Using this procedure, we obtain accurate values of the structure factor for all measured curves which are used to determine the atomic positions in the unit cell. Using the structure factors, we refined the atomic positions in the unit cell and listed their values in Table~\ref{table}. The results indicate that $P112_{1}/m$ is indeed the best fit space group with $\gamma=90.367(7)^{\circ}$.

\begin{table}[h]
\caption{\label{table} Atomic coordinates of SrIrO$_3$ films on (110)-oriented GdScO$_3$ substrates at room temperature. Space group is $P112_{1}/m$ (No. 11), $a=5.6120(5)$ {\AA}, $b=5.5865(4)$ {\AA}, $c=7.934(2)$ {\AA}, $\gamma=90.367(7)^{\circ}$.}
\begin{ruledtabular}
\begin{tabular}{ccccccc}
Atom&Wyckoff&$x$&$y$&$z$\\
\hline
Sr1 &2$e$&0.0047(2)&0.9746(11)&0.25\\
Sr2 &2$e$&0.4956(2)&0.4741(14)&0.25\\
Ir1 &2$c$&0&0.5&0\\
Ir2 &2$b$&0.5&0&0\\
O1a &2$e$&0.9420(15)&0.4999(23)&0.25\\
O1b &2$e$&0.5571(7)&0.9999(23)&0.25\\
O2a &4$f$&0.7179(23)&0.2841(7)&0.9709(12)\\
O2b &4$f$&0.7790(15)&0.7854(10)&0.9738(27)\\
\end{tabular}
\end{ruledtabular}
\end{table}

\begin{figure}\vspace{-0pt}
\includegraphics[width=8.5cm]{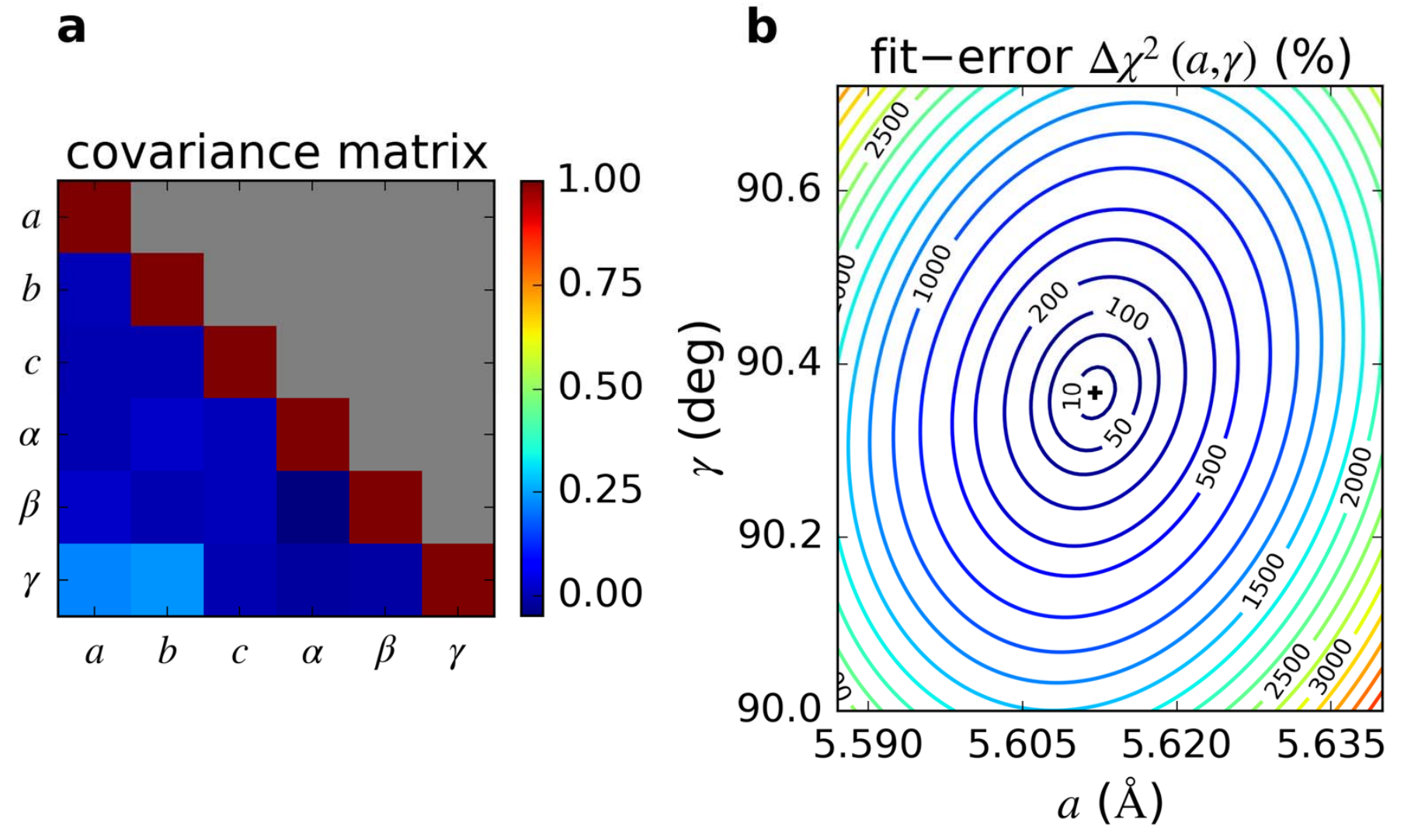}
\caption{\label{angle} (a) Covariance matrix for the unit cell parameter fit showing that all cross correlations are smaller than 0.25. (b) Relative change of the fit error as a function of the lattice parameter and monoclinic angle $\gamma$.
}
\end{figure}

To rule out fitting artifacts, we performed correlation analysis among the fitting variables, which confirms that the monoclinicity is only associated with parameters within the $ab$ plane. Specifically, the unit-cell shape parameters, i.e. the lattice parameters ($a$, $b$, $c$) and unit-cell angles ($\alpha$, $\beta$, $\gamma$), were varied in order to simultaneously fit all measured crystal truncation-rod-scans. The angles ¦Á and ¦Â were found to be 90.01(1)$^{\circ}$ and 90.00(1)$^{\circ}$, respectively. The deviation from 90$^{\circ}$ is on the order of the error bar and similar deviations were found when fitting the diffraction positions of the orthorhombic substrate indicating that our the deviations are limited by our experimental resolution. On the other hand Fig.~\ref{XRD} illustrates clearly that, in order to describe the layer peak position of selected diffraction peaks, a lowering of the symmetry to a monoclinic space group is required. To exclude possible correlations between fit parameters as the source for this monoclinicity, i.e. $\gamma\neq90^{\circ}$, we show the covariance matrix of the fit in Fig.~\ref{angle}(a). We find only values smaller than 0.25 in the covariance matrix, indicating only weak correlations between the parameters. However, the strongest correlation is just found for $\gamma$ and the unit-cell parameters, $a$ and $b$. In Fig.~\ref{angle}(b), we therefore show the relative change of the fit error as a contour plot for the parameters $\gamma$ and $a$. A clear minimum is found for the values of a = 5.6120(5) {\AA}, and ¦Ã = 90.367(7)$^{\circ}$. Under no circumstances can the structure of the film be orthorhombic, since this would cause an significant increase of the fit error by more than 1500\%.

\begin{figure}\vspace{-0pt}
\includegraphics[width=8.5cm]{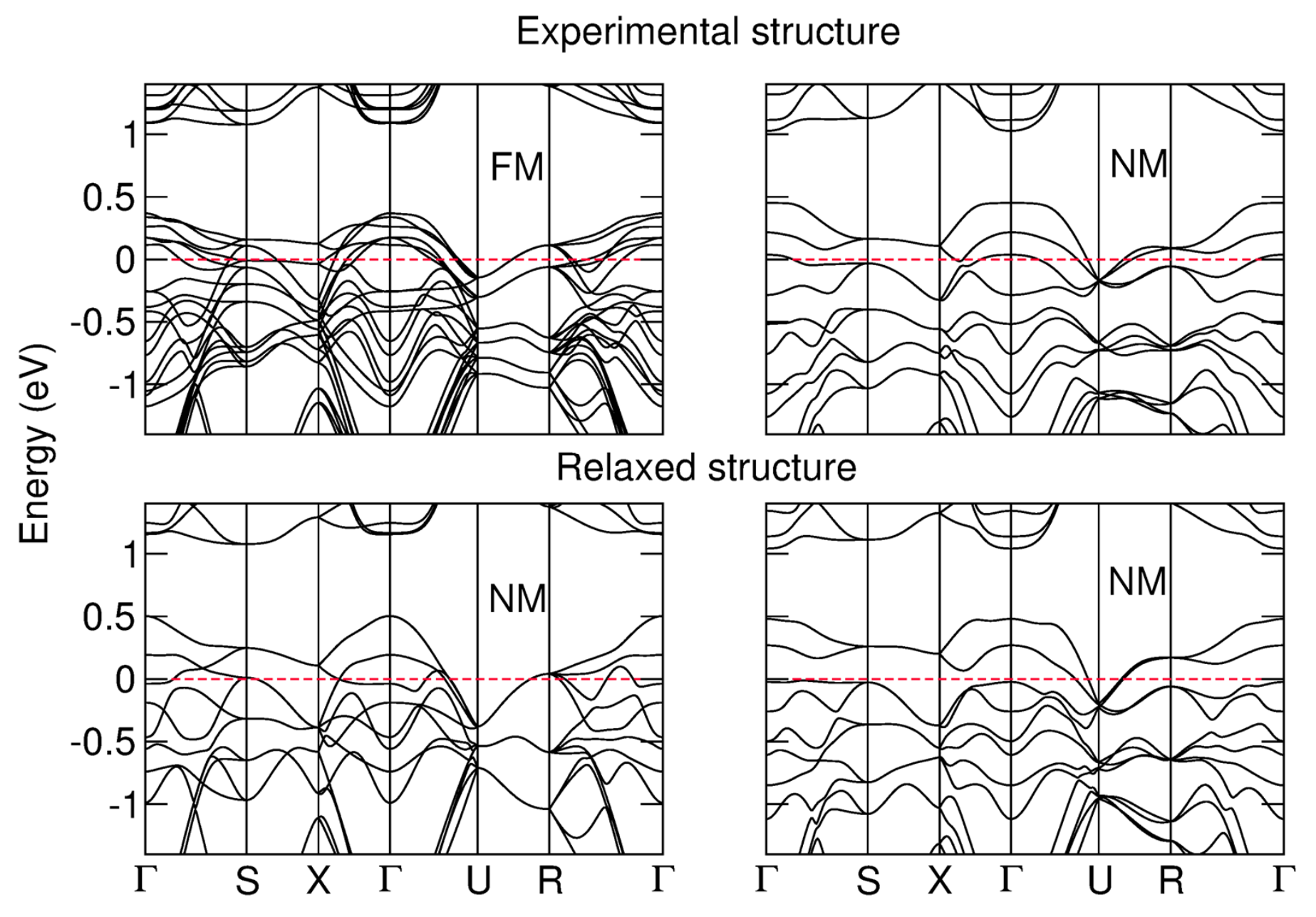}
\caption{\label{bulk} Band structures from LDA calculations without (left) and with (right) SOC. FM and NM stand for ferromagnetic metal and nonmagnetic metal, respectively.
}
\end{figure}

\textit{Density Functional Calculations}. Based on the resolved crystal structure and symmetry, we computed the band structure for our strained film by local-density approximation (LDA) calculations with SOC included. The core and valence electrons were treated using the electronic configurations $4s^{2}4p^{6}5s^{2}$ (Sr), $5d^{7}6s^{2}$ (Ir), and $2s^{2}2p^{4}$ (O); a 3$\times$3$\times$3 Monkhorst-Pack $k$-point mesh \cite{Monkhorst/Pack:1976}; and a 600~eV plane-wave cutoff. Before performing calculations for thin films under strain, we calculated the band structure for perovskite SrIrO$_3$ in bulk (strain free). Figure~\ref{bulk} shows the results from the LDA calculations without and with SOC. It can be seen that, in the absence of SOC, the degenerate $t_{2g}$ orbitals form multiple intercrossing bands, giving rise to a ferromagnetic metallic ground state with a large density of states (DOS) around the Fermi level. When including SOC splits these bands, the obtained band structure consists of the empty $e_g$-like bands found between 1 and 5 eV, and the $t_{2g}$ states split by SOC into the occupied so-called $J_{\rm eff}$=3/2 states from -0.5 to -2.5 eV and the half-filled $J_{\rm eff}$=1/2 states (0.5 to -0.4 eV). The $J_{\rm eff}=1/2$ states form four intersecting bands owing to the four-formula unit cell and band-folding. The overall band structure and dispersion are consistent with the other reported calculations for the bulk \cite{Carter2012,Zeb2012}.

\begin{figure}\vspace{-0pt}
\includegraphics[width=8.5cm]{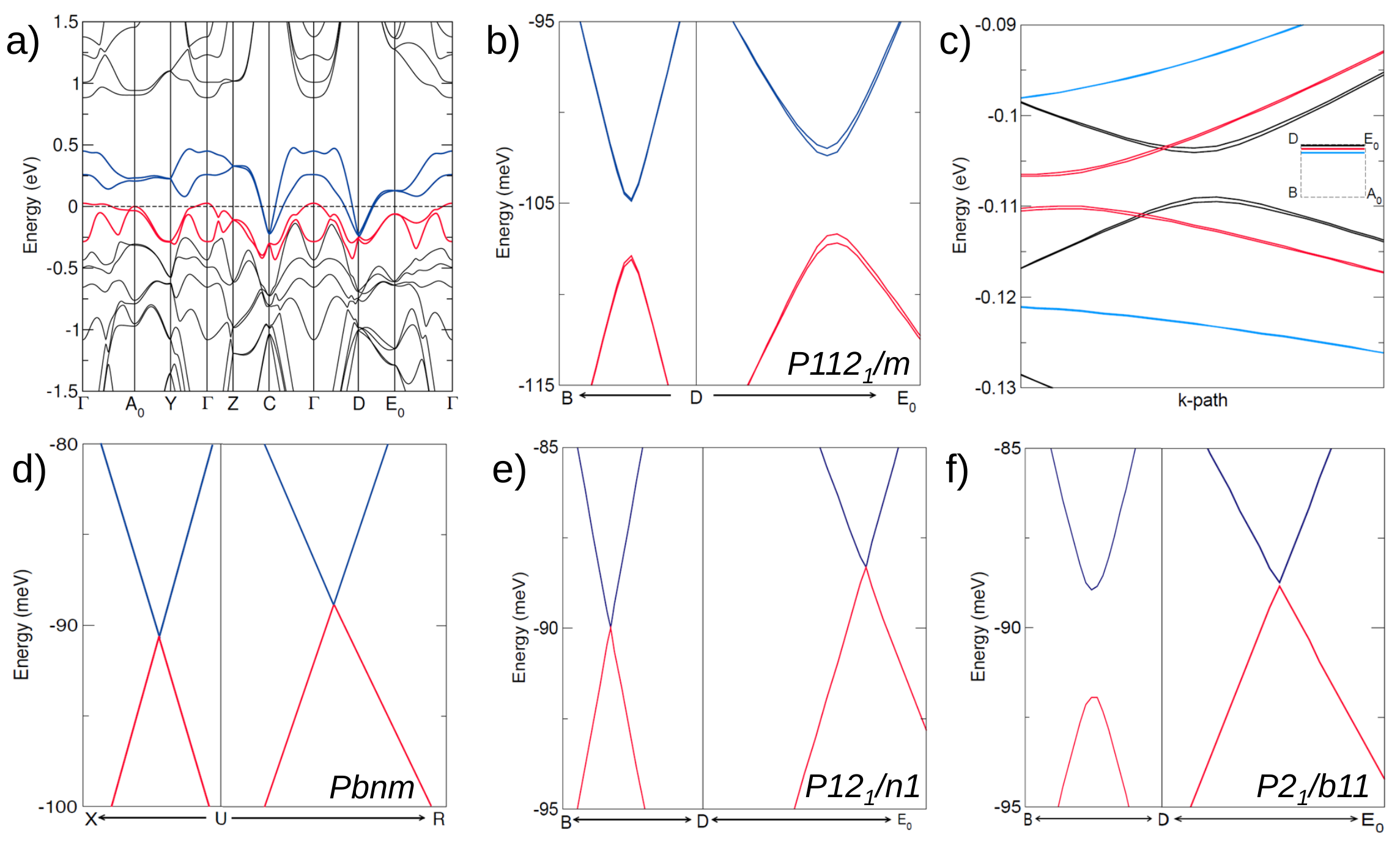}
\caption{\label{LDA}
(a) Band structures calculated for the strained film structure given in Table~\ref{table}. Red and blue bands highlight the hole-like and electron-like $J_{\rm eff}=1/2$ bands, respectively. By convention, the $U$-, $X$-, and $R$-points of the orthorhombic Brillouin zone are equivalent to the $D$-, $B$-, and $E_{o}$-points of the monoclinic one. The electronic LDA+SOC band structures near the $U$-point ($D$-point) for (b) the $P112_{1}/m$ structure (determined by the strained SrIrO$_3$ structure in Table~\ref{table}), (d) the $Pbnm$ structure of the bulk, (e) the $P12_{1}/n1$ structure obtained by removing the $b$-glide and the mirror, and (f) the $P2_{1}/b11$ structure obtained by removing the $n$-glide and the mirror. Red and blue highlight the hole-like and electron-like bands, respectively. (c) Band dispersions along three different $k$-lines (inset) in the $X-U-R$ ($B-D-E_{o}$) plane. Only results on a quarter of the $X-U-R$ plane are shown here because the other three quarters are symmetrically equivalent.
}
\end{figure}

\textit{Strain effect.} Calculation using the resolved crystal structure and symmetry of our strained film showed a similar overall band structure (Fig.~\ref{LDA}a). However, as shown in Figs.~\ref{LDA}a and b, we also found an $\sim5$ meV gap between the upper and lower Dirac cones around the $U$ point (the $D$ point under monoclinic convention), indicating lifted Dirac degeneracy. Moreover, the Dirac gap transforms the band dispersions near the original Dirac nodes from linear to quadratic, i.e. a larger particle mass. To ensure the entire nodal ring is gapped, we surveyed multiple low-symmetry $k$-lines near the $U$ point and confirmed that no point node exists (Fig.~\ref{LDA}c). This gap was unexpected because the Dirac line nodes were suggested to be a robust feature originating from the mirror symmetry \cite{Carter2012}, which we found to be the only symmetry operation left in the film $P112_{1}/m$ structure from the strain-free $Pbnm$ space group (in addition to inversion). Note that the resolved structure of our film is the only experimental input to our calculation. To further validate this result, we examined our calculations on the bulk $Pbnm$ structure (Fig.~\ref{bulk}) and confirmed that the linear band crossing on the Dirac nodal ring is indeed well preserved therein (Fig.~\ref{LDA}d).

\textit{Symmetry analysis}. To reconcile this contradiction and understand the symmetry protection on the Dirac nodal ring, we compared the symmetry breaking in our film to that of the Sr$_{2}$IrRhO$_{6}$ superlattice structure previously proposed for lifting the Dirac degeneracy \cite{Carter2012}. Note that, in the proposed superlattice, the mirror symmetry is designed to be broken by alternating the perovskite $B$-sites along the $Pbnm$ $c$ axis with Ir and Rh. Since this $B$-site alternation is only along the $c$ axis, the $b$-glide operation within the $ab$ plane is preserved. However, the $n$-glide which involves a $c$-axis translation is required be broken simultaneously with the mirror symmetry. Indeed, the mass term introduced to achieve the mirror symmetry breaking also anticommutes with the $n$-glide operator \cite{Carter2012}. The resulting space group of a superlattice of this kind would be $P2_{1}/b11$. Compared to our strained $P112_{1}/m$ structure, it is clear that the common broken symmetry is the $n$-glide. Therefore, to account for the lifted Dirac degeneracy in both structures, one can infer that the Dirac line nodes around the $U$ point are protected by the $n$-glide symmetry operation, in addition to inversion and time reversal.

We further corroborated this conclusion by performing calculations on structures of the $P12_{1}/n1$ and $P2_{1}/b11$ space groups, where only the $n$-glide and the $b$-glide are left, respectively. The results showed that the Dirac nodal ring is indeed fully preserved under $P12_{1}/n1$ (Fig.~\ref{LDA}e), demonstrating that the mirror or the $b$-glide is not essential in the protection. Indeed, the nodal ring is also gapped under $P2_{1}/b11$ (Fig.~\ref{LDA}f) except for a pair of Dirac point nodes along the $U-R$ ($D-E_{0}$) line. This is consistent with the calculation in Ref.~\cite{Carter2012} where the nodal ring becomes nodal points in the Sr$_{2}$IrRhO$_{6}$ superlattice. The difference is that the symmetry-removal was achieved here by unit-cell distortion rather than superlattice construction. Nevertheless, this observation shows that, in the absence of the $n$-glide, the Dirac points on this high-symmetry $k$-line can be protected by the remaining $b$-glide symmetry. This is also consistent with the case of $P112_{1}/m$ discussed above; when both $n$-glide and $b$-glide are removed, no Dirac points remain, and the nodal ring is fully gapped.

\section{IV. Discussion}
This result leads to several direct consequences and implications for understanding orthoperovskite iridates. First of all, one must consider the epitaxy-symmetry interplay when studying thin film samples. Because strain is inevitable in epitaxial films, the crystal symmetry might be lowered and the Dirac degeneracy might be lifted. It is conceivable that the Dirac gap size will depend on the degree of deviation from the $n$-glide symmetry and increase with strain. Examples of strained SrIrO$_{3}$ films of this kind include that used in recent ARPES studies reported by Y. F. Nie $et$ $al.$ \cite{Nie2015} and Liu $et$ $al.$ \cite{Liu2015}. The former revealed the semimetallic bands with Dirac-cone like electron pockets, and the later reported a gap of 30-50 meV between the electron-like and hole-like cones near the $U$ point with suppressed linearity, consistent with our conclusion. However, the epitaxy-induced symmetry-breaking was not considered in either study. A description of the band structure taking into account the structure and symmetry is clearly requisite.

Due to the protection by the $n$-glide operation, the Dirac semimetallic band crossing falls into the nonsymmorphic category of three-dimensional Dirac semimetals, which is now known as a fundamental class of topological semimetals \cite{Young2012,Young2015}. While other known nonsymmorphic Dirac semimetals are characterized by point nodes \cite{Young2012,Steinberg2014}, SrIrO$_3$ is distinctive with its Dirac line node. Additionally, as discussed above, partial symmetry lowering may tune it into point nodes protected by another nonsymmorphic symmetry, the $b$-glide operation. In fact, recently, we became aware of another theoretical work by Fang $et$ $al.$ \cite{Fang}, which also indicates that a four-band crossing nodal line on a $k=\pi$ plane, such as that in SrIrO$_3$, can be protected by the nonsymmorphic symmetry in the presence of SOC.

We can also further deduce the situations of other epitaxial orientations, which we categorized here based on pseudo-cubic orientations. Due to the nature of a glide plane, removing it without complex superlattice structures requires strain to distort the reflection axis, i.e. changing $\alpha$ and/or $\gamma$ for the $n$-glide operation. The (100)$_{\rm c}$-orientation has three $Pbnm$ variants, (110), (1$\overline{1}$0), and (001). It is obvious that (1$\overline{1}$0) will induce the same monoclinic distortion and symmetry-breaking as (110). But the $n$-glide is preserved in the case of (001) because [001] is collinear with the primitive axis and all angles may remain 90$^{\circ}$ except oblique surface lattices, e.g. LaAlO$_{3}$. The same is true for the (100)- and (010)-orientations, two of the (110)$_{\rm c}$ variants. The third variant, the (111)-orientation, is, however, opposite since the symmetry would become triclinic as strain is applied along all primitive directions. Finally, (101) and (011) are the two (111)$_{\rm c}$ variants relevant to many proposals of topological insulators in SrIrO$_{3}$-based heterostructures \cite{Xiao2011}. For (101), since the $b$-axis is within the surface and remains normal to the $ac$ plane, the $n$-glide is preserved. It is, however, broken for (011) as this orthogonality is lost, i.e. $\alpha\neq90^{\circ}$. The symmetry will decrease to $P2_{1}/b11$, gapping the nodal ring except for a pair of point nodes (Fig.~\ref{LDA}f). Therefore, the Dirac degeneracy removal sensitively depends on the epitaxial orientation. Whether Dirac gaps of different sizes can be open under the same amount of strain but different orientations would be an interesting subject for further study.

Finally, when the Dirac gap opens, the SOC-correlation interplay may significantly change as well. Note that Zeb and Kee \cite{Zeb2012} discovered from their calculations of bulk SrIrO$_3$ that the larger SOC is, the weaker the correlation strength is, in sharp contrast to the insulating iridates, e.g. Sr$_2$IrO$_4$, where SOC is believed to assist correlation \cite{Kim2008}. This unusual behavior in SrIrO$_3$ is caused by the existence of the Dirac line nodes, which prevents full charge gap opening despite moderately strong correlation. However, the situation may change if the Dirac degeneracy is lifted; when the Dirac cones are free to renormalize under correlation, the SOC-correlation interplay may become synergistic. Therefore, while strain and film thickness may be used to engineer the bandwidth, electronic, and transport properties, whether the $n$-glide symmetry and Dirac degeneracy are broken could lead to different responses. Recent resistivity measurements by several groups indeed reported strain- and thickness-dependent behavior and metal-insulator transitions in SrIrO$_{3}$ films \cite{Wu2010,Biswas2008,Gruenewald2014,Zhang2015}. However, consensus on the underlying origin has not been reached, and the epitaxy-induced symmetry breaking remains to be completely taken into account.

\section{VI. Conclusions}
In conclusion, by investigating the structural and electronic change in epitaxially strained SrIrO$_{3}$ films, we showed that the epitaxy-induced symmetry breaking alone is sufficient to remove the Dirac degeneracy of the band crossing around the $U$ point. We identified the symmetry operation protecting the Dirac degeneracy as the $n$-glide operation, which was hidden in the previous theoretical proposal for SrIrO$_{3}$-based topological insulators \cite{Carter2012}. This finding places the system in the nonsymmorphic class of Dirac semimetals and provides new avenues to open the Dirac gap in much simpler strained-film structures and potentially stabilizing topologically insulating phases. Relieving this symmetry-protected band crossing may also reverse the SOC-correlation interplay. Our study thus highlights the critical role of epitaxy-induced symmetry-breaking, which must be taken into account in understanding and designing topological materials.

\section{Acknowledgments}
\begin{acknowledgments}
We thank C. Fang, P. J. Ryan and J.-W. Kim for insightful discussions. We thank J. Rubio for experiment assistance. J.L. is sponsored by the Science Alliance Joint Directed Research and Development Program at the University of Tennessee. We are thankful for support from the Director, Office of Science, Office of Basic Energy Sciences, Materials Sciences and Engineering Division, of the U.S. Department of Energy under Contract No. DE-AC02-05CH11231. We acknowledge additional support of the material synthesis facility through the D.O.D. ARO MURI, E3S, and DARPA. D.K. acknowledges the support by the Austrian Science Fund (FWF): J3523-N27. We acknowledge support from the Grant Agency of the Czech Republic Grant no. 14-37427. Financial support from the Spanish MINECO (MAT2012-33207) is acknowledged. We acknowledge ESRF for the provision of beamtime. Use of the Advanced Photon Source, an Office of Science User Facility operated for the U.S. Department of Energy (DOE) Office of Science by Argonne National Laboratory, was supported by the U.S. DOE under Contract No. DE-AC02-06CH11357. J.M.R. and D.P. acknowledge support from the Army Research Office under Grant No. W911NF-15-1-0017 and the High Performance Computing Modernization Program (HPCMP) of the DOD for providing computational resources that have contributed to the research results reported herein.
\end{acknowledgments}

\end{document}